\documentclass[pra,aps,showpacs,amsmath,amssymb,twocolumn]{revtex4}

\usepackage{graphicx}
\usepackage{dcolumn}
\usepackage{bbm}

\newcommand{\be}{\begin{equation}}
\newcommand{\ee}{\end{equation}}
\newcommand{\ba}{\begin{array}}
\newcommand{\ea}{\end{array}}
\newcommand{\bea}{\begin{eqnarray}}
\newcommand{\eea}{\end{eqnarray}}

\newcommand{\bra}[1]{\ensuremath{\langle #1 |}}
\newcommand{\ket}[1]{\ensuremath{| #1 \rangle}}

\begin{document}

\title{Stabilising entanglement by quantum jump-based feedback}

\author{A. R. R. Carvalho}
\affiliation{Department of Physics, Faculty of Science, The Australian National University, ACT 0200, Australia}
\author{J. J. Hope}
\affiliation{Australian Centre for Quantum-Atom Optics, Department of Physics, Faculty of Science, The Australian National University, ACT 0200, Australia}

\date{\today}

\begin{abstract}
We show that direct feedback based on quantum jump detection can be used to generate entangled steady states. We present a strategy that is insensitive to detection inefficiencies and robust against errors in the control Hamiltonian. This feedback procedure is also shown to overcome spontaneous emission effects by stabilising states with high degree of entanglement.

\end{abstract}


\maketitle

Numerous applications of quantum information theory require the ability to produce entangled states and perform controlled operations on them. There have been many successful experiments in this direction~\cite{bouwmeester99,rauschenbeutel00,sackett00, panPRL01, roos} but despite the effort to screen the system against unwanted imperfections and interactions, entanglement degradation through uncontrolled coupling with the environment remains a major obstacle~\cite{eberly_04,arrc_mpd}. Even if experiments were to be performed under perfect conditions, fundamental factors such as spontaneous emission in atomic qubits~\cite{roosPRL04} would persist, limiting the lifetime of entangled states and demanding efficient schemes to protect them. 

Recent experimental developments have enabled real-time monitoring and manipulation of individual quantum systems~\cite{geremia04,reiner04,morrow02,bushev06}, suggesting that quantum feedback control~\cite{belav,wise_milb93,wiseman94,doherty99}, may emerge as a natural possible route to develop strategies to prepare entangled states and prevent their deterioration. Recent attempts have been made in this direction with proposals to control both continuous~\cite{mancini06,mancini06b} and discrete~\cite{stockton04,mancini05,wang05} variable entanglement, using either Bayesian~\cite{doherty99} or Markovian (direct)~\cite{wise_milb93,wiseman94} feedback scheme. While in the latter strategy a simple feedback directly proportional to the detection signal is used, in the former, control depends on an estimate of the system state based on the information acquired from the measurement results. 
Although this can result in an improvement over the direct feedback scheme, it also comes at the cost of an increasing complexity in the experimental implementation due to the (challenging) need for a real time estimation of the quantum state.

In this Letter we will show that a Markovian feedback scheme  based on the continuous monitoring of quantum jumps, together with  an appropriate choice for the feedback Hamiltonian, can lead to an improvement, in amount and robustness, of the steady state entanglement in a model of two driven and collectively damped qubits~\cite{schneider02,wang05}. In the absence of spontaneous emission, a pure maximally entangled state is dynamically generated irrespective of detection inefficiencies. Furthermore, this strategy is also able to cope with spontaneous emission effects by stabilising highly entangled states.
 
\begin{figure}
\includegraphics[width=4.5cm]{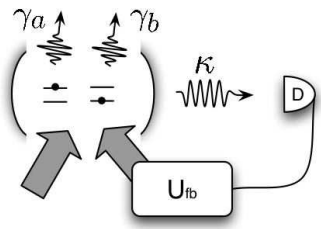}
\caption{Schematic view of the model. The system consists of a pair of two-level simultaneously driven by a laser and coupled to a damped cavity. Conditioned on the measurement of the output of the leaky cavity, a Hamiltonian is applied to the atoms, completing the feedback scheme.} 
\label{fig1}
\end{figure}

 Our system consists of a pair of two-level atoms equally, and resonantly, coupled to a single cavity mode, with a coupling strength $g$. The atoms can spontaneously decay
with rates $\gamma_1$ and $\gamma_2$, and are simultaneously driven by a laser
field (see Fig.~\ref{fig1}). The cavity mode is damped,
and in the limit where its decay rate $\kappa$ is very large, it can be
adiabatically eliminated leading to the following
master equation for the atomic degrees of freedom~\cite{wang05}:
\be\label{eq:me_total}
\dot \rho=-i \Omega \left[(J_+ + J_-),\rho \right] + \Gamma {\cal D}[J_-]\rho + \gamma_1{\cal D}[\sigma_1]\rho + \gamma_2 {\cal D}[\sigma_2]\rho .
\ee
Here, $\Omega$ is the effective Rabi frequency of the collective driving, $\Gamma=g^2/\kappa$ is the collective decay rate of the
atoms and the cavity, and superoperator $\cal D$ acting on an operator $c$ is given by $ {\cal D}[c]\rho \equiv c \rho c^{\dagger}-\left(c^{\dagger}c \rho+\rho c^{\dagger}c\right)/2$.
The Hamiltonian is written in terms of angular momentum operators $J_{-}=\sigma_1+\sigma_2$ and $J_+=\sigma_1^{+}+\sigma_2^{+}$, where $\sigma_i=\ket{g_i}\bra{e_i}$ and $\sigma_i^{+}=\ket{e_i}\bra{g_i}$ are,
respectively, the lowering and raising operators for the $i$-th two level atom.
In the limit that the collective decay rate is much larger than the
spontaneous emission rates, $\Gamma \gg \gamma_1,\, \gamma_2$, we recover the Dicke model~\cite{agarwal74}
\be
\label{eq:me_dicke}
\dot \rho= {\cal L} \rho=-i\Omega\left[(J_+ + J_-),\rho \right] + \Gamma {\cal D}[J_-]\rho.
\ee
Before investigating the influence of feedback in this equation, let us first briefly analyse the entanglement properties of its steady state solutions~\cite{schneider02}. The first important feature of Eq.~(\ref{eq:me_dicke}) is  that it is symmetric with respect to exchange of the atoms. This suggests that, instead of using the two qubit basis $\{ \ket{gg},\,\ket{ge},\,\ket{eg},\,\ket{ee}\}$, one should use angular momentum
states, and analyse the system in terms of the symmetric ($j=1$)
\be
\label{symm}
\ket{1}=\ket{gg}, \:\:\: \ket{2}=\frac{\ket{ge}+\ket{eg}}{\sqrt{2}}, \:\:\: \ket{3}=\ket{ee},
\ee
and anti-symmetric ($j=0$)
\be
\label{asymm}
\ket{4}=\frac{\ket{ge}-\ket{eg}}{\sqrt{2}}
\ee
subspaces.

A simple inspection shows that $\ket{4}$ is a stationary state solution of Eq.~(\ref{eq:me_dicke}). Despite the triviality of the dynamics in this subspace, it shouldn't be regarded as a totally uninteresting case as far as entanglement production is concerned, since its asymptotic state is a pure, maximally entangled one. In fact, this situation was explored in a recent proposal for producing Werner states in a system of atoms inside a cavity~\cite{agarwal06}. 
On the other hand, in the symmetric subspace, entanglement is dynamically generated from any initial condition, even from initially separable states. However, even for optimal parameters, the amount of entanglement (given by the concurrence~\cite{wot98}) in this case is only about $10 \%$ of the Bell state's value~\cite{schneider02}.

Now we can complete the scenario by introducing the description of the measurement scheme and feedback. The idea is depicted in Fig.~\ref{fig1}: the cavity output is measured by a photo-detector $\rm D$ whose signal provides the input to the application of the control Hamiltonian $F$. Note that, in this kind of monitoring, the absence of signal predominates and the control is only triggered after a detection click, {\it i.e.} a quantum jump, occurs. The unconditioned master equation for this case was derived by Wiseman in~\cite{wiseman94} and, for our system, it reads 
\bea
\label{eq:pd}
\dot \rho &=&-i \Omega\left[(J_+ + J_-),\rho \right] + \Gamma {\cal D}[U_{\rm fb} J_-]\rho. 
\eea
The jump feedback is easier to interpret when one write the last term of Eq.~\ref{eq:pd} explicitly: ${\cal D}[U_{\rm fb} J_-]= U_{\textrm{fb}} J_- \rho J_+ U_{\textrm{fb}}^{\dagger}-(J_+ J_- \rho+\rho J_+ J_-)/2$. The unitary transformation $U_{\rm fb}=\exp\left[-i F \delta t/\hbar \right]$, representing the finite amount of evolution imposed by the control Hamiltonian on the system, is only applied immediately after a detection (or jump) event, which is described by the action of $J_-$ in the first term of the superoperator ${\cal D}[J_-]$. Note that the anti-symmetric Bell state $\ket{4}$ remains a stationary solution independently of the form of $U_{\rm fb}$.

Once the measurement prescription has been chosen, the freedom to design a feedback to produce a steady state with the desired properties lies in the different choices for the feedback operator $U_{\rm fb}$. Although this represents an enormous range of possibilities, even when considering the limitations imposed by  experimental constraints, here we will restrict to two different cases. 

\begin{figure}
\includegraphics[width=8.0cm]{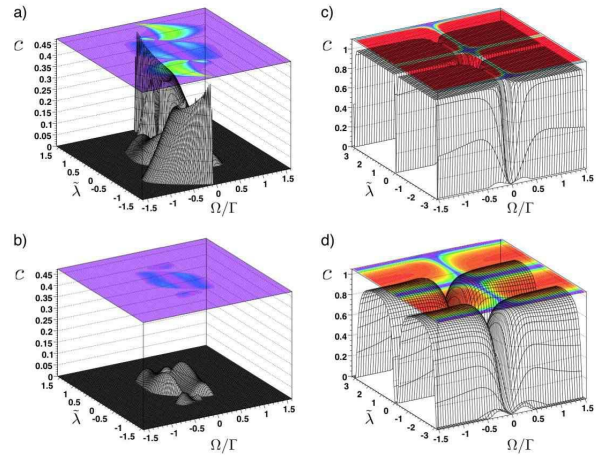}
\caption{(Color online) Steady state concurrence as a function of feedback and driving strengths for $J_x$ (left) and $\sigma_x$ (right) control, with (bottom) or without (top) spontaneous emission effects. While the influence of decay is pronounced in the $J_x$ control, the local strategy is left basically unaltered. In this case, and with $\gamma/\Gamma=0.01$, a highly entangled state ($c=0.95$) is stabilised for almost all parameters. 
} 
\label{figure3d}
\end{figure}

Our first choice is the feedback $U_{\rm fb}= \exp[-i \tilde \lambda J_x]$ that preserves the symmetry properties with respect to exchange of atoms. This coincides with the driving Hamiltonian in Eq.~(\ref{eq:me_dicke}), and was used in~\cite{wang05} to show that a feedback control based on homodyne detection can increase the asymptotic entanglement by 3 times as compared to the non-controlled case. Although this is a substantial augment, it corresponds to only $30 \%$ of the maximum possible value for the entanglement. 
In the jump-based feedback, the steady state solutions in the symmetric subspace can be analytically calculated from Eq.~(\ref{eq:pd}) and the corresponding entanglement are shown as a function of the driving and feedback strengths in Fig.~\ref{figure3d}a. The maximum concurrence, $c\approx 0.45$, occurs at $\Omega/\Gamma \approx 0.08$ and $\tilde \lambda \approx \pm -1.49$, exceeding the value obtained via homodyne-based feedback. Despite this further improvement, the absolute value of the entanglement provided by this feedback strategy is still far from the maximum value $c=1$. Moreover, its performance will worsen considerably when spontaneous emission is added to Eq.~(\ref{eq:me_dicke}), as shown in Fig.~\ref{figure3d}a for $\gamma_1=\gamma_2\equiv \gamma=0.01 \, \Gamma$. Even with such a small value, atomic decay plays a crucial role here: it breaks the symmetry of the system, opening the way for interference between different subspaces to occur. This is the explanation of the disappearance of the higher peaks from Fig.~\ref{figure3d}a to Fig.~\ref{figure3d}b: the corresponding stationary solution shows a large component of the state $\ket{2}$ that interferes destructively with the state $\ket{4}$, which is driven by the anti-symmetric part of the dynamics.

An alternative would be to choose a feedback that, like the spontaneous emission term, does not preserve the symmetry with respect to exchange of atoms. This may allow the feedback to move population between the subspaces, limiting the possibilities of destructive interference.  Fortunately, as we shall show, this hope is realised with the simplest choice, yet experimentally realistic~\cite{nussmann_05}, of a symmetry-breaking Hamiltonian, where the control acts on just one of the atoms. We call this {\it local feedback} and represent its action as $U_{\rm fb}=U_1 \otimes \mathbb{I}$. 
One can now replace this form in Eq.~(\ref{eq:me_dicke}), project it in the basis defined by Eqs.~(\ref{symm}) and~(\ref{asymm}) using a general unitary $U_1$, and then set $\dot \rho=0$ to find the stationary states. The system of $16$ equations obtained from this procedure admits a single solution, namely the anti-symmetric Bell state $\ket{4}$. Therefore, opposed to the uncontrolled case where this state has to be produced beforehand and is then unaffected by the dynamics, now a pure maximally entangled state is dynamically generated for all initial conditions and parameters (excluding the trivial cases of absence of feedback or driving).  We illustrate this in Fig.~\ref{figure3d}c for the particular choice $U_{1}= \exp[-i \tilde \lambda \sigma_x]$, with $\sigma_x=\sigma_1+\sigma_1^\dag$.

\begin{figure}
\includegraphics[width=6.5cm]{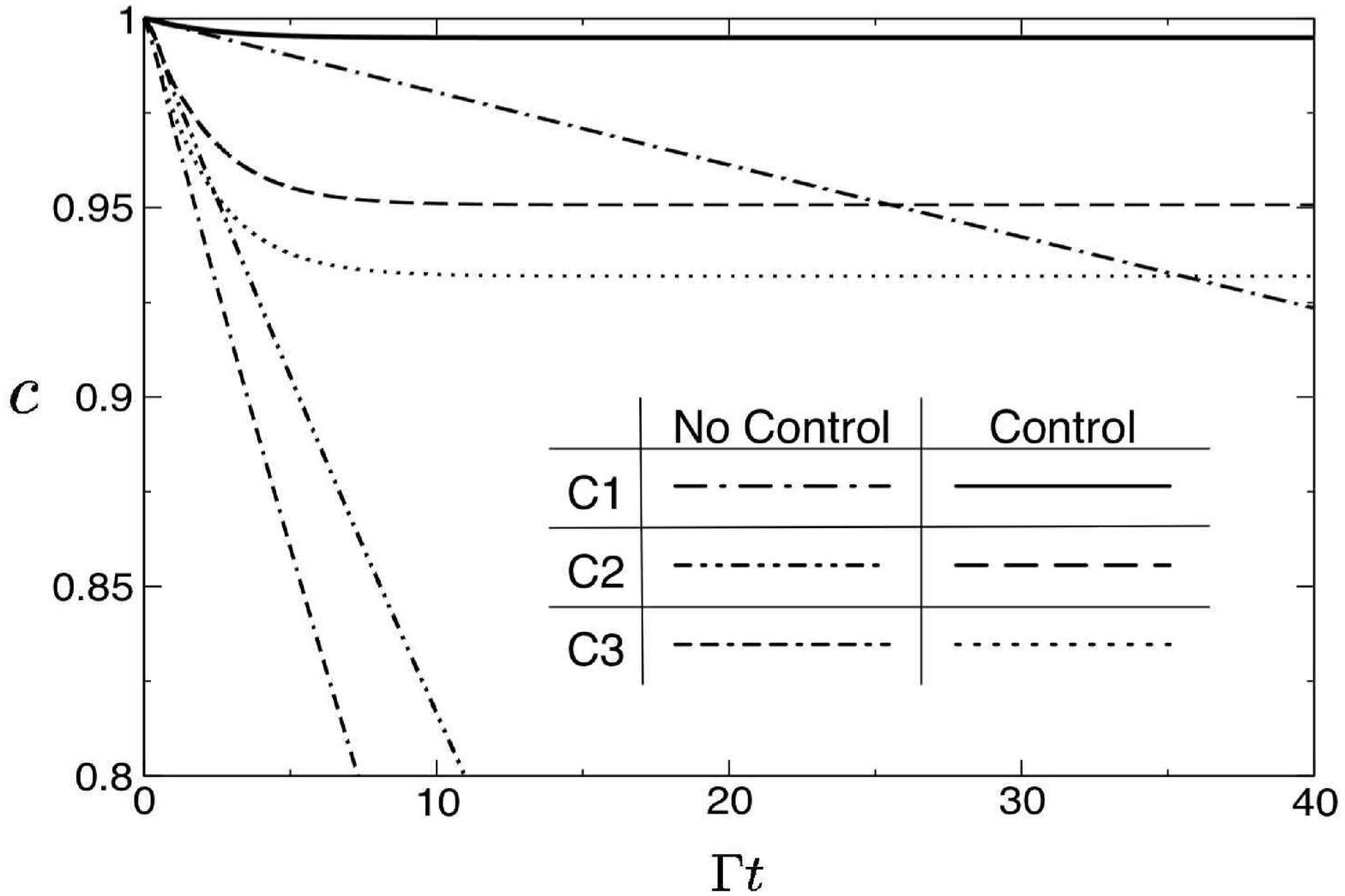}
\caption{Time evolution of concurrence for $\Omega=3\Gamma$ and initial state $\ket{4}$ in the presence of spontaneous emission effects for the cases where $\gamma=0.001 \, \Gamma$ ($\textrm{C}_1$) and $\gamma= 0.01\, \Gamma$ ($\textrm{C}_2$). Remarkably, the local $\sigma_x$ feedback control with $\tilde \lambda=\pi/2$ is able to stabilise the state with large amount of entanglement even when an extra dephasing is added ($\gamma_{\textrm{deph}}=\gamma=0.01\,\Gamma$ in $\textrm{C}_3$).
} 
\label{figSE}
\end{figure}

But it is only when spontaneous emission is taken into account that the advantage of {\it jump-based} local feedback turns to be really remarkable. A comparison between the time evolution of entanglement for an initial anti-symmetric Bell state in the controlled and non-controlled cases in the presence of atomic decay, and $U_1$ as defined above, is shown in Fig.~\ref{figSE}. Spontaneous emission takes the system away from the anti-symmetric subspace and the dynamics without feedback is not able to restore the Bell state $\ket{4}$ (cases $\textrm{C}_1$ and $\textrm{C}_2$ for $\gamma/\Gamma=0.001$ and $0.01$, respectively). Instead, decay terms increase the symmetric component as they move the system to its ground state while the other terms will tend to drive it to the steady state of Eq.~(\ref{eq:me_dicke}). Conversely, in the presence of feedback, the system evolves under two competing dynamics: while the feedback pushes the state $\ket{gg}$ to $\ket{4}$, spontaneous decay forces the system the other way around. The final steady state entanglement is, therefore, set by the balance between those two opposing tendencies, decreasing exponentially with the ratio $\gamma/\Gamma$. Remarkably, the proposed control also works when different decoherence sources are considered: in case $\textrm{C}_3$ of Fig.~\ref{figSE}, besides spontaneous emission, extra local dephasing terms $\gamma_{\textrm{deph}}{\cal D}[\sigma_i^{\dag} \sigma_i]$ were added and a high value of steady state entanglement is still obtained for $\gamma_{\textrm{deph}}=\gamma=0.01\,\Gamma$.
Consequently, as soon as the ratio between decoherence and collective decay rates remain small, the final entangled state will be very close to the anti-symetric Bell state as shown by the control curves in Fig.~\ref{figSE}. For spontaneous emission rates ($\textrm{C}_1$ and $\textrm{C}_2$) consistent with recent experimental parameters~\cite{boozer06}, concurrence remains above $c\approx 0.95$. Note that this performance is only possible with the combination of a local control {\it and} jump-based feedback. If the same control was used under a homodyne-based feedback, the maximum attainable entanglement would be $c\approx 0.72$ ($\gamma=0$) and $c\approx 0.60$ ($\gamma=0.01\,\Gamma$), for $\Omega/\Gamma \approx 0.18$ and $\tilde \lambda\approx \pm -0.12$~\cite{arrc_control}.

The behaviour as a function of $\tilde \lambda$ and $\Omega$ for this choice of feedback Hamiltonian remains basically unchanged (Fig.~\ref{figure3d}d), as compared with Fig.~\ref{figure3d}c, with only  the values of stationary entanglement slightly reduced. This is again remarkable, as errors or fluctuations in the feedback and driving strengths would not affect significantly the final entanglement.
In the absence of decay, the steady state was independent of the form of $U_1$, provided it was not the identity, which corresponds to the absence of feedback.  However, the \textit{rate} at which feedback induced that steady state was affected by the choice of control.  As the high entanglement in the presence of decay is due to a competition between the decay and the feedback rate, it is not surprising that different choices of $U_1$ lead to different steady state values.  Fortunately, these variations are very small, particularly for perturbations around our choice of $\sigma_x$.   Thus our scheme will be robust against imperfections in the feedback Hamiltonian. 

\begin{figure}
\includegraphics[width=7.0cm]{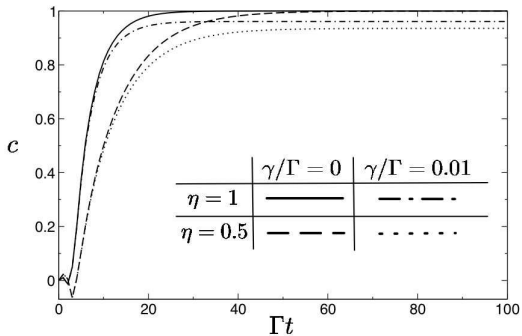}
\caption{Effect of different detection efficiencies on entanglement evolution for a $\sigma_x$ control with $\Omega=0.4 \Gamma$, $\tilde \lambda=\pi/2$, and both atoms initially in the ground state. Without decay the system always reaches the anti-symmetric Bell state but at times depending on the efficiency, while with spontaneous emission there is a small decrease in the final entanglement.
} 
\label{figeta}
\end{figure}

Up to now, our analysis has neglected effects of inefficiencies in the detection process, which may be important, as feedback relies on the manipulation of the system based on information gained by the measurement. In the jump feedback case, the extension of Eq.~(\ref{eq:pd}) to allow for a inefficient detection can be done by identifying two distinct situations when a jump occurs: in the first the detector clicks and the feedback transformation $U_{\rm fb}$ is applied, in the second the detector fails to click and there is no control action. The corresponding equation reads 
\be
\label{eq:pd_eta}
\dot \rho=-\frac{i}{\hbar}\Omega\left[(J_+ + J_-),\rho \right] + \Gamma \eta {\cal D}[U_{\textrm{fb}} J_-]\rho + \Gamma \left(1-\eta\right) {\cal D}[J_-]\rho.
\ee
When the detector efficiency $\eta$ is zero, no information is extracted from the measurement and the equation reduces to Eq.~(\ref{eq:me_dicke}) where no feedback is applied. Evidently, in the limit of perfect detection Eq.~(\ref{eq:pd}) is regained, and, for a local control, and without spontaneous decay, a maximally entangled steady state is reached. In the intermediate case where $0<\eta<1$, one would expect that imperfect knowledge gain should lead to a worse control. However, the anti-symmetric Bell state is a steady state of both Eqs.~(\ref{eq:me_dicke}) and~(\ref{eq:pd}), and one can show, proceeding exactly as in the unit-efficiency case, that this also holds true for Eq.~(\ref{eq:pd_eta}) for any $\eta >0$. The effect is illustrated in Fig.~\ref{figeta} where the time evolution of concurrence is shown for a fixed driving frequency $\Omega=0.4 \Gamma$, and for the $\sigma_x$ feedback Hamiltonian with $\tilde \lambda=\pi/2$.
Without atomic decay, a non-unit detection efficiency ($\eta=0.5$) simply delays the time at which stationarity is achieved as compared to the $\eta=1$ case.  In the presence of atomic decay there will be also a decrease in the asymptotic entanglement as $\eta$ decreases. This should be also expected as less efficient detection will represent an effectively weaker control and therefore a change in the balance between decay and feedback dynamics.
However, in the limit considered here ($\Gamma/\gamma \gg 1$) this decrease is small (see Fig.~\ref{figeta}). 

In conclusion, we have proposed a quantum jump based feedback scheme to prepare and stabilise highly entangled states in the presence of spontaneous emission and decoherence. By monitoring the environment and feeding back with a suitable interaction, in our case a local one, we were able to modify the master equation in such a way that its steady state coincides with the target state. The strategy works when this engineered dynamics possesses a single stationary state and is much stronger than the undesired decay effects~\cite{arrc_sp}. The scheme performs well in the presence of detection inefficiencies and is also robust against imperfections in the preparation of the feedback Hamiltonian. 

Evidently, many different aspects of the problem remain open. A comparison between feedback schemes using different monitoring methods and controls is under investigation and will be presented elsewhere~\cite{arrc_control}. Also relevant, from the point of view of scalability requirements in quantum information, would be the extension to higher number of atoms. Would this strategy also be efficient for a larger number of them? Even in the case of two atoms the possibilities are vast and the question whether the present results can be further improved by designing better feedback schemes is open. 

\bibliography{qcontrol}

\begin{thebibliography}{29}
\expandafter\ifx\csname natexlab\endcsname\relax\def\natexlab#1{#1}\fi
\expandafter\ifx\csname bibnamefont\endcsname\relax
  \def\bibnamefont#1{#1}\fi
\expandafter\ifx\csname bibfnamefont\endcsname\relax
  \def\bibfnamefont#1{#1}\fi
\expandafter\ifx\csname citenamefont\endcsname\relax
  \def\citenamefont#1{#1}\fi
\expandafter\ifx\csname url\endcsname\relax
  \def\url#1{\texttt{#1}}\fi
\expandafter\ifx\csname urlprefix\endcsname\relax\def\urlprefix{URL }\fi
\providecommand{\bibinfo}[2]{#2}
\providecommand{\eprint}[2][]{\url{#2}}

\bibitem[{\citenamefont{Bouwmeester et~al.}(1999)\citenamefont{Bouwmeester,
  Pan, Daniell, Weinfurter, and Zeilinger}}]{bouwmeester99}
\bibinfo{author}{\bibfnamefont{D.}~\bibnamefont{Bouwmeester}},
  \bibinfo{author}{\bibfnamefont{J.-W.} \bibnamefont{Pan}},
  \bibinfo{author}{\bibfnamefont{M.}~\bibnamefont{Daniell}},
  \bibinfo{author}{\bibfnamefont{H.}~\bibnamefont{Weinfurter}},
  \bibnamefont{and}
  \bibinfo{author}{\bibfnamefont{A.}~\bibnamefont{Zeilinger}},
  \bibinfo{journal}{Phys. Rev. Lett.} \textbf{\bibinfo{volume}{82}},
  \bibinfo{pages}{1345} (\bibinfo{year}{1999}).

\bibitem[{\citenamefont{Rauschenbeutel
  et~al.}(2000)\citenamefont{Rauschenbeutel, Nogues, Osnaghi, Bertet, Brune,
  Raimond, and Haroche}}]{rauschenbeutel00}
\bibinfo{author}{\bibfnamefont{A.}~\bibnamefont{Rauschenbeutel}},
  \bibinfo{author}{\bibfnamefont{G.}~\bibnamefont{Nogues}},
  \bibinfo{author}{\bibfnamefont{S.}~\bibnamefont{Osnaghi}},
  \bibinfo{author}{\bibfnamefont{P.}~\bibnamefont{Bertet}},
  \bibinfo{author}{\bibfnamefont{M.}~\bibnamefont{Brune}},
  \bibinfo{author}{\bibfnamefont{J.-M.} \bibnamefont{Raimond}},
  \bibnamefont{and} \bibinfo{author}{\bibfnamefont{S.}~\bibnamefont{Haroche}},
  \bibinfo{journal}{Science} \textbf{\bibinfo{volume}{288}},
  \bibinfo{pages}{2024} (\bibinfo{year}{2000}).

\bibitem[{\citenamefont{Sackett et~al.}(2000)\citenamefont{Sackett, Kielpinski,
  King, Langer, Meyer, Myatt, Rowe, Turchette, Itano, Wineland
  et~al.}}]{sackett00}
\bibinfo{author}{\bibfnamefont{C.~A.} \bibnamefont{Sackett}},
  \bibinfo{author}{\bibfnamefont{D.}~\bibnamefont{Kielpinski}},
  \bibinfo{author}{\bibfnamefont{B.~E.} \bibnamefont{King}},
  \bibinfo{author}{\bibfnamefont{C.}~\bibnamefont{Langer}},
  \bibinfo{author}{\bibfnamefont{V.}~\bibnamefont{Meyer}},
  \bibinfo{author}{\bibfnamefont{C.~J.} \bibnamefont{Myatt}},
  \bibinfo{author}{\bibfnamefont{M.}~\bibnamefont{Rowe}},
  \bibinfo{author}{\bibfnamefont{Q.~A.} \bibnamefont{Turchette}},
  \bibinfo{author}{\bibfnamefont{W.~M.} \bibnamefont{Itano}},
  \bibinfo{author}{\bibfnamefont{D.~J.} \bibnamefont{Wineland}},
  \bibnamefont{et~al.}, \bibinfo{journal}{Nature}
  \textbf{\bibinfo{volume}{404}}, \bibinfo{pages}{256} (\bibinfo{year}{2000}).

\bibitem[{\citenamefont{Pan et~al.}(2001)\citenamefont{Pan, Daniell, Gasparoni,
  Weihs, and Zeilinger}}]{panPRL01}
\bibinfo{author}{\bibfnamefont{J.-W.} \bibnamefont{Pan}},
  \bibinfo{author}{\bibfnamefont{M.}~\bibnamefont{Daniell}},
  \bibinfo{author}{\bibfnamefont{S.}~\bibnamefont{Gasparoni}},
  \bibinfo{author}{\bibfnamefont{G.}~\bibnamefont{Weihs}}, \bibnamefont{and}
  \bibinfo{author}{\bibfnamefont{A.}~\bibnamefont{Zeilinger}},
  \bibinfo{journal}{Phys. Rev. Lett.} \textbf{\bibinfo{volume}{86}},
  \bibinfo{pages}{4435} (\bibinfo{year}{2001}).

\bibitem[{\citenamefont{Roos et~al.}(2004{\natexlab{a}})\citenamefont{Roos,
  Riebe, H\"affner, H\"ansel, elm, Lancaster, Becher, Schmidt-Kaler, and
  Blatt}}]{roos}
\bibinfo{author}{\bibfnamefont{C.~F.} \bibnamefont{Roos}},
  \bibinfo{author}{\bibfnamefont{M.}~\bibnamefont{Riebe}},
  \bibinfo{author}{\bibfnamefont{H.}~\bibnamefont{H\"affner}},
  \bibinfo{author}{\bibfnamefont{W.}~\bibnamefont{H\"ansel}},
  \bibinfo{author}{\bibfnamefont{J.~B.} \bibnamefont{elm}},
  \bibinfo{author}{\bibfnamefont{G.~P.~T.} \bibnamefont{Lancaster}},
  \bibinfo{author}{\bibfnamefont{C.}~\bibnamefont{Becher}},
  \bibinfo{author}{\bibfnamefont{F.}~\bibnamefont{Schmidt-Kaler}},
  \bibnamefont{and} \bibinfo{author}{\bibfnamefont{R.}~\bibnamefont{Blatt}},
  \bibinfo{journal}{Science} \textbf{\bibinfo{volume}{304}},
  \bibinfo{pages}{1478} (\bibinfo{year}{2004}{\natexlab{a}}).

\bibitem[{\citenamefont{Yu and Eberly}(2004)}]{eberly_04}
\bibinfo{author}{\bibfnamefont{T.}~\bibnamefont{Yu}} \bibnamefont{and}
  \bibinfo{author}{\bibfnamefont{J.~H.} \bibnamefont{Eberly}},
  \bibinfo{journal}{Phys. Rev. Lett.} \textbf{\bibinfo{volume}{93}},
  \bibinfo{pages}{140404} (\bibinfo{year}{2004}).

\bibitem[{\citenamefont{Carvalho et~al.}(2004)\citenamefont{Carvalho, Mintert,
  and Buchleitner}}]{arrc_mpd}
\bibinfo{author}{\bibfnamefont{A.~R.~R.} \bibnamefont{Carvalho}},
  \bibinfo{author}{\bibfnamefont{F.}~\bibnamefont{Mintert}}, \bibnamefont{and}
  \bibinfo{author}{\bibfnamefont{A.}~\bibnamefont{Buchleitner}},
  \bibinfo{journal}{Phys. Rev. Lett.} \textbf{\bibinfo{volume}{93}},
  \bibinfo{pages}{230501} (\bibinfo{year}{2004}).

\bibitem[{\citenamefont{Roos et~al.}(2004{\natexlab{b}})\citenamefont{Roos,
  Lancaster, Riebe, H\"affner, H\"ansel, Gulde, Becher, Eschner, Schmidt-Kaler,
  and Blatt}}]{roosPRL04}
\bibinfo{author}{\bibfnamefont{C.~F.} \bibnamefont{Roos}},
  \bibinfo{author}{\bibfnamefont{G.~P.~T.} \bibnamefont{Lancaster}},
  \bibinfo{author}{\bibfnamefont{M.}~\bibnamefont{Riebe}},
  \bibinfo{author}{\bibfnamefont{H.}~\bibnamefont{H\"affner}},
  \bibinfo{author}{\bibfnamefont{W.}~\bibnamefont{H\"ansel}},
  \bibinfo{author}{\bibfnamefont{S.}~\bibnamefont{Gulde}},
  \bibinfo{author}{\bibfnamefont{C.}~\bibnamefont{Becher}},
  \bibinfo{author}{\bibfnamefont{J.}~\bibnamefont{Eschner}},
  \bibinfo{author}{\bibfnamefont{F.}~\bibnamefont{Schmidt-Kaler}},
  \bibnamefont{and} \bibinfo{author}{\bibfnamefont{R.}~\bibnamefont{Blatt}},
  \bibinfo{journal}{Phys. Rev. Lett.} \textbf{\bibinfo{volume}{92}},
  \bibinfo{pages}{220402} (\bibinfo{year}{2004}{\natexlab{b}}).

\bibitem[{\citenamefont{Geremia et~al.}(2004)\citenamefont{Geremia, Stockton,
  and Mabuchi}}]{geremia04}
\bibinfo{author}{\bibfnamefont{J.~M.} \bibnamefont{Geremia}},
  \bibinfo{author}{\bibfnamefont{J.~K.} \bibnamefont{Stockton}},
  \bibnamefont{and} \bibinfo{author}{\bibfnamefont{H.}~\bibnamefont{Mabuchi}},
  \bibinfo{journal}{Science} \textbf{\bibinfo{volume}{304}},
  \bibinfo{pages}{270} (\bibinfo{year}{2004}).

\bibitem[{\citenamefont{Reiner et~al.}(2004)\citenamefont{Reiner, Smith,
  Orozco, Wiseman, and Gambetta}}]{reiner04}
\bibinfo{author}{\bibfnamefont{J.~E.} \bibnamefont{Reiner}},
  \bibinfo{author}{\bibfnamefont{W.~P.} \bibnamefont{Smith}},
  \bibinfo{author}{\bibfnamefont{L.~A.} \bibnamefont{Orozco}},
  \bibinfo{author}{\bibfnamefont{H.~M.} \bibnamefont{Wiseman}},
  \bibnamefont{and} \bibinfo{author}{\bibfnamefont{J.}~\bibnamefont{Gambetta}},
  \bibinfo{journal}{Phys. Rev. A} \textbf{\bibinfo{volume}{70}},
  \bibinfo{pages}{023819} (\bibinfo{year}{2004}).

\bibitem[{\citenamefont{Morrow et~al.}(2002)\citenamefont{Morrow, Dutta, and
  Raithel}}]{morrow02}
\bibinfo{author}{\bibfnamefont{N.~V.} \bibnamefont{Morrow}},
  \bibinfo{author}{\bibfnamefont{S.~K.} \bibnamefont{Dutta}}, \bibnamefont{and}
  \bibinfo{author}{\bibfnamefont{G.}~\bibnamefont{Raithel}},
  \bibinfo{journal}{Phys. Rev. Lett.} \textbf{\bibinfo{volume}{88}},
  \bibinfo{pages}{093003} (\bibinfo{year}{2002}).

\bibitem[{\citenamefont{Bushev et~al.}(2006)\citenamefont{Bushev, Rotter,
  Wilson, Dubin, Becher, Eschner, Blatt, Steixner, Rabl, and
  Zoller}}]{bushev06}
\bibinfo{author}{\bibfnamefont{P.}~\bibnamefont{Bushev}},
  \bibinfo{author}{\bibfnamefont{D.}~\bibnamefont{Rotter}},
  \bibinfo{author}{\bibfnamefont{A.}~\bibnamefont{Wilson}},
  \bibinfo{author}{\bibfnamefont{F.}~\bibnamefont{Dubin}},
  \bibinfo{author}{\bibfnamefont{C.}~\bibnamefont{Becher}},
  \bibinfo{author}{\bibfnamefont{J.}~\bibnamefont{Eschner}},
  \bibinfo{author}{\bibfnamefont{R.}~\bibnamefont{Blatt}},
  \bibinfo{author}{\bibfnamefont{V.}~\bibnamefont{Steixner}},
  \bibinfo{author}{\bibfnamefont{P.}~\bibnamefont{Rabl}}, \bibnamefont{and}
  \bibinfo{author}{\bibfnamefont{P.}~\bibnamefont{Zoller}},
  \bibinfo{journal}{Phys. Rev. Lett.} \textbf{\bibinfo{volume}{96}},
  \bibinfo{pages}{043003} (\bibinfo{year}{2006}).

\bibitem[{\citenamefont{Belavkin}(1999)}]{belav}
\bibinfo{author}{\bibfnamefont{V.~P.} \bibnamefont{Belavkin}},
  \bibinfo{journal}{Rep. Math. Phys.} \textbf{\bibinfo{volume}{43}},
  \bibinfo{pages}{405} (\bibinfo{year}{1999}).

\bibitem[{\citenamefont{Wiseman and Milburn}(1993)}]{wise_milb93}
\bibinfo{author}{\bibfnamefont{H.~M.} \bibnamefont{Wiseman}} \bibnamefont{and}
  \bibinfo{author}{\bibfnamefont{G.~J.} \bibnamefont{Milburn}},
  \bibinfo{journal}{Phys. Rev. Lett.} \textbf{\bibinfo{volume}{70}},
  \bibinfo{pages}{548} (\bibinfo{year}{1993}).

\bibitem[{\citenamefont{Wiseman}(1994)}]{wiseman94}
\bibinfo{author}{\bibfnamefont{H.~M.} \bibnamefont{Wiseman}},
  \bibinfo{journal}{Phys. Rev. A} \textbf{\bibinfo{volume}{49}},
  \bibinfo{pages}{2133} (\bibinfo{year}{1994}).

\bibitem[{\citenamefont{Doherty and Jacobs}(1999)}]{doherty99}
\bibinfo{author}{\bibfnamefont{A.~C.} \bibnamefont{Doherty}} \bibnamefont{and}
  \bibinfo{author}{\bibfnamefont{K.}~\bibnamefont{Jacobs}},
  \bibinfo{journal}{Phys. Rev. A} \textbf{\bibinfo{volume}{60}},
  \bibinfo{pages}{2700} (\bibinfo{year}{1999}).

\bibitem[{\citenamefont{Mancini}(2006)}]{mancini06}
\bibinfo{author}{\bibfnamefont{S.}~\bibnamefont{Mancini}},
  \bibinfo{journal}{Phys. Rev. A} \textbf{\bibinfo{volume}{73}},
  \bibinfo{pages}{010304(R)} (\bibinfo{year}{2006}).

\bibitem[{\citenamefont{Mancini and Wiseman}(2006)}]{mancini06b}
\bibinfo{author}{\bibfnamefont{S.}~\bibnamefont{Mancini}} \bibnamefont{and}
  \bibinfo{author}{\bibfnamefont{H.~M.} \bibnamefont{Wiseman}}
  (\bibinfo{year}{2006}), \bibinfo{note}{preprint arXiv quant-ph/0610006}.

\bibitem[{\citenamefont{Stockton et~al.}(2004)\citenamefont{Stockton, van
  Handel, and Mabuchi}}]{stockton04}
\bibinfo{author}{\bibfnamefont{J.~K.} \bibnamefont{Stockton}},
  \bibinfo{author}{\bibfnamefont{R.}~\bibnamefont{van Handel}},
  \bibnamefont{and} \bibinfo{author}{\bibfnamefont{H.}~\bibnamefont{Mabuchi}},
  \bibinfo{journal}{Phys. Rev. A} \textbf{\bibinfo{volume}{70}},
  \bibinfo{pages}{022106} (\bibinfo{year}{2004}).

\bibitem[{\citenamefont{Mancini and Wang}(2005)}]{mancini05}
\bibinfo{author}{\bibfnamefont{S.}~\bibnamefont{Mancini}} \bibnamefont{and}
  \bibinfo{author}{\bibfnamefont{J.}~\bibnamefont{Wang}},
  \bibinfo{journal}{Eur. Phys. J. D} \textbf{\bibinfo{volume}{32}},
  \bibinfo{pages}{257} (\bibinfo{year}{2005}).

\bibitem[{\citenamefont{Wang et~al.}(2005)\citenamefont{Wang, Wiseman, and
  Milburn}}]{wang05}
\bibinfo{author}{\bibfnamefont{J.}~\bibnamefont{Wang}},
  \bibinfo{author}{\bibfnamefont{H.~M.} \bibnamefont{Wiseman}},
  \bibnamefont{and} \bibinfo{author}{\bibfnamefont{G.~J.}
  \bibnamefont{Milburn}}, \bibinfo{journal}{Phys. Rev. A}
  \textbf{\bibinfo{volume}{71}}, \bibinfo{pages}{042309}
  (\bibinfo{year}{2005}).

\bibitem[{\citenamefont{Schneider and Milburn}(2002)}]{schneider02}
\bibinfo{author}{\bibfnamefont{S.}~\bibnamefont{Schneider}} \bibnamefont{and}
  \bibinfo{author}{\bibfnamefont{G.~J.} \bibnamefont{Milburn}},
  \bibinfo{journal}{Phys. Rev. A} \textbf{\bibinfo{volume}{65}},
  \bibinfo{pages}{042107} (\bibinfo{year}{2002}).

\bibitem[{\citenamefont{Agarwal}(1974)}]{agarwal74}
\bibinfo{author}{\bibfnamefont{G.~S.} \bibnamefont{Agarwal}},
  \bibinfo{journal}{Springer Tracs Mod. Phys.} \textbf{\bibinfo{volume}{70}},
  \bibinfo{pages}{1} (\bibinfo{year}{1974}).

\bibitem[{\citenamefont{Agarwal and Kapale}(2006)}]{agarwal06}
\bibinfo{author}{\bibfnamefont{G.~S.} \bibnamefont{Agarwal}} \bibnamefont{and}
  \bibinfo{author}{\bibfnamefont{K.~T.} \bibnamefont{Kapale}},
  \bibinfo{journal}{Phys. Rev. A} \textbf{\bibinfo{volume}{73}},
  \bibinfo{pages}{022315} (\bibinfo{year}{2006}).

\bibitem[{\citenamefont{Wootters}(1998)}]{wot98}
\bibinfo{author}{\bibfnamefont{W.~K.} \bibnamefont{Wootters}},
  \bibinfo{journal}{Phys. Rev. Lett.} \textbf{\bibinfo{volume}{80}},
  \bibinfo{pages}{2245} (\bibinfo{year}{1998}).

\bibitem[{\citenamefont{Nu{\ss}mann et~al.}(2005)\citenamefont{Nu{\ss}mann,
  Hijlkema, Weber, Rohde, Rempe, and Kuhn}}]{nussmann_05}
\bibinfo{author}{\bibfnamefont{S.}~\bibnamefont{Nu{\ss}mann}},
  \bibinfo{author}{\bibfnamefont{M.}~\bibnamefont{Hijlkema}},
  \bibinfo{author}{\bibfnamefont{B.}~\bibnamefont{Weber}},
  \bibinfo{author}{\bibfnamefont{F.}~\bibnamefont{Rohde}},
  \bibinfo{author}{\bibfnamefont{G.}~\bibnamefont{Rempe}}, \bibnamefont{and}
  \bibinfo{author}{\bibfnamefont{A.}~\bibnamefont{Kuhn}},
  \bibinfo{journal}{Phys. Rev. Lett.} \textbf{\bibinfo{volume}{95}},
  \bibinfo{pages}{173602} (\bibinfo{year}{2005}).

\bibitem[{\citenamefont{Boozer et~al.}(2006)\citenamefont{Boozer, Boca, Miller,
  Northup, and Kimble}}]{boozer06}
\bibinfo{author}{\bibfnamefont{A.~D.} \bibnamefont{Boozer}},
  \bibinfo{author}{\bibfnamefont{A.}~\bibnamefont{Boca}},
  \bibinfo{author}{\bibfnamefont{R.}~\bibnamefont{Miller}},
  \bibinfo{author}{\bibfnamefont{T.~E.} \bibnamefont{Northup}},
  \bibnamefont{and} \bibinfo{author}{\bibfnamefont{H.~J.}
  \bibnamefont{Kimble}}, \bibinfo{journal}{Phys. Rev. Lett.}
  \textbf{\bibinfo{volume}{97}}, \bibinfo{pages}{083602}
  (\bibinfo{year}{2006}).

\bibitem[{\citenamefont{Carvalho and Hope}(2006)}]{arrc_control}
\bibinfo{author}{\bibfnamefont{A.~R.~R.} \bibnamefont{Carvalho}}
  \bibnamefont{and} \bibinfo{author}{\bibfnamefont{J.~J.} \bibnamefont{Hope}}
  (\bibinfo{year}{2006}), \bibinfo{note}{in preparation}.

\bibitem[{\citenamefont{Carvalho et~al.}(2001)\citenamefont{Carvalho, Milman,
  de~Matos~Filho, and Davidovich}}]{arrc_sp}
\bibinfo{author}{\bibfnamefont{A.~R.~R.} \bibnamefont{Carvalho}},
  \bibinfo{author}{\bibfnamefont{P.}~\bibnamefont{Milman}},
  \bibinfo{author}{\bibfnamefont{R.~L.} \bibnamefont{de~Matos~Filho}},
  \bibnamefont{and}
  \bibinfo{author}{\bibfnamefont{L.}~\bibnamefont{Davidovich}},
  \bibinfo{journal}{Phys. Rev. Lett.} \textbf{\bibinfo{volume}{86}},
  \bibinfo{pages}{4988} (\bibinfo{year}{2001}).

\end{thebibliography}

\end{document}